\documentclass[article,twocolumn,amsmath,floats,showpacs,nofootinbib]{revtex4}
\usepackage[usenames]{color}

\usepackage{graphicx}
\usepackage{epstopdf}
\usepackage{textcomp}
\usepackage{hyperref}
\usepackage{multirow}

\hypersetup{colorlinks=true}

\begin{document}

\title{Spectroscopy of phonons in dirty superconducting contacts}

\author{I.K. Yanson, N.L. Bobrov, L.F. Rybal'chenko, and V.V. Fisun}
\affiliation{B.I.~Verkin Institute for Low Temperature Physics and
Engineering, of the National Academy of Sciences
of Ukraine, prospekt Lenina, 47, Kharkov 61103, Ukraine and A. M. Gor'kov State University, Khar'kov
Email address: bobrov@ilt.kharkov.ua}
\published {(\href{http://fntr.ilt.kharkov.ua/fnt/pdf/9/9-11/f09-1155r.pdf}{Fiz. Nizk. Temp.}, \textbf{9}, 1155 (1983)); (Sov. J. Low Temp. Phys., \textbf{9}, 596 (1983)}
\date{\today}

\begin{abstract}The nonlinearities of current-voltage characteristics (IVC) of superconducting niobium point contacts for voltages of the order of the characteristic phonon frequencies are investigated. It is shown that in limiting by dirty contacts ${{l}_{1}}\ll d$ , (${{l}_{1}}$  is the momentum mean-free path of electrons), features are observed on the IVC corresponding to maxima in the first or second derivatives and are situated at energies close to the characteristic phonon energies. These features are due to the energy dependence of the excess current, since the nonlinearity of the IVC in the normal state is several times smaller in absolute magnitude, and its second derivative is monotonic. For a contact diameter $d$ of the order of several tens of angstroms (the contact resistance is of the order of 100 $\Omega $), the thermal effects can be neglected. It is assumed that in the region of the constriction, the characteristic lengths of the system satisfy the relation $\xi \gtrsim d>{{\Lambda }_{\varepsilon }}$ (${{\Lambda }_{\varepsilon }}$ is the diffusion length for energy relaxation of the electrons, $\xi$  is the coherence length) which makes possible inelastic transitions of electrons between pair chemical potentials differing in energy by an account $eV$ ($V$ is the voltage drop across the contact).

\end{abstract}

\maketitle

\section{INTRODUCTION}
Point-contact spectroscopy (PCS) of the electron-phonon interaction (EPI) in normal metals permits measuring directly the spectral function of the EPI if the mean-free path of electrons is greater than the size of the contact and the temperature is sufficiently low ($kT\ll \hbar {{\omega }_{\max }}$, where $\hbar {{\omega }_{\max }}$ is limiting phonon frequency). If the metal is superconducting at low temperatures, then for investigations using the PCS method, it is transformed into the normal state with a magnetic field exceeding the critical value ${{H}_{c}}$. For superconductors with not too large values of ${{H}_{c}}$ (for example, $Pb, Sn, In, Tl, Nb$, etc.), measurement of the point contact spectra does not present any difficulties. However, it is not easy to destroy superconductivity of intermetallic compounds with a structure of the A15 type, especially if we take into account the fact that the deformation of the metal, arising with the creation of a contact in liquid helium, causes the required fields to exceed appreciably the ${{H}_{c}}$ of the undeformed specimen. For these materials, the characteristic frequencies of the phonon spectrum are likewise usually quite low, so that ${{H}_{c}}$ likewise cannot be decreased by increasing the temperature, without decreasing too much the resolution of the method.

As was first shown in Ref. \cite{Khotkevich}, the IVC of clean $S-c-S$ contacts are nonlinear and, in addition, features (maxima in $dV/dI$) are observed in their derivatives, situated at characteristic energies of the phonon spectrum and arising due to the energy dependence of the excess current. An attempt to explain the decrease in the excess current by equilibrium heating of the metal in the region of the contact, undertaken in Ref. \cite{Khotkevich}, was only partially successful, inasmuch as it did not make it possible to determine the positions of the features on the $V$ axis, which did not coincide with the positions of the peaks in the EPI function.

It was shown in recently published theoretical papers (Refs. \cite{Khlus1,Khlus2} that for clean $S-c-S$ and $S-c-N$ contacts, the IVC is nonlinear and, in addition, for $eV\gg \Delta $ ($\Delta $ is the energy gap), it can be represented as a sum of four terms
\begin{equation}
\label{eq__1}
{I(V)=\frac{V}{{{R}_{0}}}+I_{N}^{(1)}(eV)+I_{exc}^{(0)}+I_{S}^{(1)}(eV),}
\end{equation}
where $I_{N}^{(1)}(eV)$  is the negative increment to the current, coinciding with the nonlinear part of the IVC in the normal state and having a magnitude of the order of $dV/{{R}_{0}}l(eV)$. The second derivative of $I_{N}^{(1)}(eV)$ with respect to the voltage is proportional to the point contact EPI function ${{g}_{pc}}(\omega )$  (Ref. \cite{Yanson}). In Eq. \ref{eq__1}, the voltage independent excess current $I_{exc}^{(0)}$  for $T\ll \Delta $ , equals $4\Delta /3e{{R}_{0}}$ and $8\Delta /3e{{R}_{0}}$ for $S-c-N$ and $S-c-S$ contacts, respectively, while the energy dependent part of the excess current  which $I_{S}^{(1)}(eV)$ is of the order of $dI_{exc}^{(0)}/l(eV)$  and for bias voltages $eV\gg \Delta $  represents a small (of the order of $\Delta /eV$ ) increment to the nonlinear part of the IVC. An experimental investigation of IVC of clean $S-c-N$ contacts by Kamarchuk et al.\cite{Kamarchuk} confirmed the results of this theory. It showed that the second derivatives of the IVC of $Cu-Sn$ heterocontacts in the normal and superconducting states practically coincide, if the feature appearing due to the breakdown of the superconducting state in $Sn$ due to injection of quasiparticles from $Cu$ are overlooked. In the $S-c-N$ contacts investigated in Ref. \cite{Kamarchuk}, these features, as a rule, occurred beyond the limits of the phonon spectrum and their amplitude was small. They practically did not interfere with the observation of the nonlinearities due to the second and fourth terms in the sum (\ref{eq__1}). We recall that in clean tin $S-c-S$ contacts \cite{Khotkevich} nonequilibrium effects of the suppression of superconductivity on the edges are so large that against their background, the theoretically predicted \cite{Khlus1} small non-linearities due to EPI are not visible. Thus the study of point contact spectra of clean $S-c-N$ heterocontacts could make it possible to reconstruct the EPI function of superconductors with high critical parameters, without transforming them into the normal state. However, the spectra measured in this case represent a superposition of the EPI functions of the superconducting and normal metals \cite{Shekhter}, and it is difficult to separate the contribution of each metal from them.

Another problem arises in connection with the fact that in the process of creating the contact, a high concentration of elastic scatterers of electrons (impurities or structural defects) is introduced into the microconstriction, although they are absent in a massive specimen. For this reason, most of the point contacts obtained in the experiment correspond more to the dirty limit than the clean limit.

The purpose of this work is to investigate experimentally the nonlinearities of IVC of limiting dirty point contacts with identical superconducting electrodes in the region of bias voltages $eV$, corresponding to the characteristic energies of the phonon spectrum ($eV\gg \Delta $). Niobium is chosen as the metal because, on the one hand, the oxidized surface does not contain a large density of centers that strongly scatter electrons and, on the other, its phonon spectrum and EPI function have been well studied experimentally \cite{Robinson,Wolf}. Niobium contacts can be easily switched into the normal state with a magnetic field.

We discovered that in cases when heating of the metal in the microconstriction can be neglected, the nonlinear IVC has features (maxima in the first or second derivatives) situated at bias voltages corresponding to the characteristic phonon energies. These features are due to the energy dependence of the excess current, and the related nonlinearity of the IVC is several times greater than the non-linearity of the IVC in the normal state. We also note that the derivatives of the IVC in the normal state do not contain maxima. In other words, the EPI spectrum can be observed in the superconducting state when it is absent in the normal state.

\section{PROCEDURE FOR OBTAINING POINT CONTACTS}

Niobium is a material that is used in most works concerning the experimental investigation of the properties of superconducting point contacts. However, in all cases, the structure and purity of the metal in the region of the constriction remained undetermined. Meanwhile, it is known that the electrical properties of niobium oxides, which always cover the electrodes, vary over a wide range depending on the composition: from superconducting $NbO\text{ }\left( {{T}_{c}}=1.4\  K \right)$  to the semiconductor $N{{b}_{2}}{{O}_{5}}$. A higher concentration of depairing centers, suppressing superconductivity near the surface, is possible in thin niobium oxide layers. Point-contact spectroscopy first permitted obtaining direct information on the structure and composition of the metal in the region of the constriction. In our work, this information is used to construct an adequate model of the contact.

All of the specimens investigated can be separated into two groups, depending on the method used to obtain them, in which the barrier, which is opaque to electrons, separating the electrodes was created either from niobium oxides or by total oxidation of a thin layer of aluminum deposited on clean surfaces of niobium electrodes in a high vacuum. The characteristics of both types of contacts were similar, with the exception that the percentage yield of contacts of the second type suitable for measurements was higher.

We used single-crystalline $Nb$ with a resistance ratio ${{\rho }_{300}}/{{\rho }_{10}}=100$  as the material for the electrodes. We cut out $2\times 2\times 12\ m{{m}^{3}}$  electrodes using an electroerosion method. Then we chemically etched them in a mixture of concentrated acids $HF:HCl{{O}_{4}}:HN{{O}_{3}}$ , used with equal volume ratios, washed the specimens in distilled water, dried them, and mounted them into a holder. In some experiments (contacts belonging to the second group), after etching, we placed the electrodes into a vacuum chamber and heated them with a sharply focused electron beam gun ($U=4\div 5\text{ }kV,\text{ }I=5\div 10\text{ }mA$) to a pre melting temperature over a period of 5-7 min under a pressure of $(1\div 2)\cdot {{10}^{-7}}mm\ Hg$. After cooling to room temperature, we deposited an aluminum layer 100-150 ${\AA}$  thick on the faces of the electrodes. We oxidized the aluminum in a boiling $30\%$ solution of ${{H}_{2}}{{O}_{2}}$.

We created the point contacts using the shear technique \cite{Yanson}. The holder with a torsion-type damper permitted regulating the clamping force holding the electrodes to one another over a wide range. In the first contacts, the IVC often had a semiconducting nature, i.e., the differential resistance decreased with increasing voltage. An increase of the clamping force usually led to metallic conductivity in the region of the constriction.

The current-voltage characteristics and their derivatives were regulated with the help of a standard modulation technique. For measurements in the normal state, the superconductivity of the electrodes was suppressed by a magnetic field with an intensity of up to 50 $kOe$ at a temperature of $4.2~K$. Insertion and removal of the magnetic field destroyed the contacts, so that we performed the measurements of the characteristics for the same contact in the superconducting and normal state in helium vapor after increasing the temperature from 4.3 to $10~K$.

\section{EXPERIMENTAL RESULTS AND DISCUSSION}
\subsection{Point contact spectroscopy of the electron-phonon interaction in niobium in the normal state}
\begin{figure}[]
\includegraphics[width=8cm,angle=0]{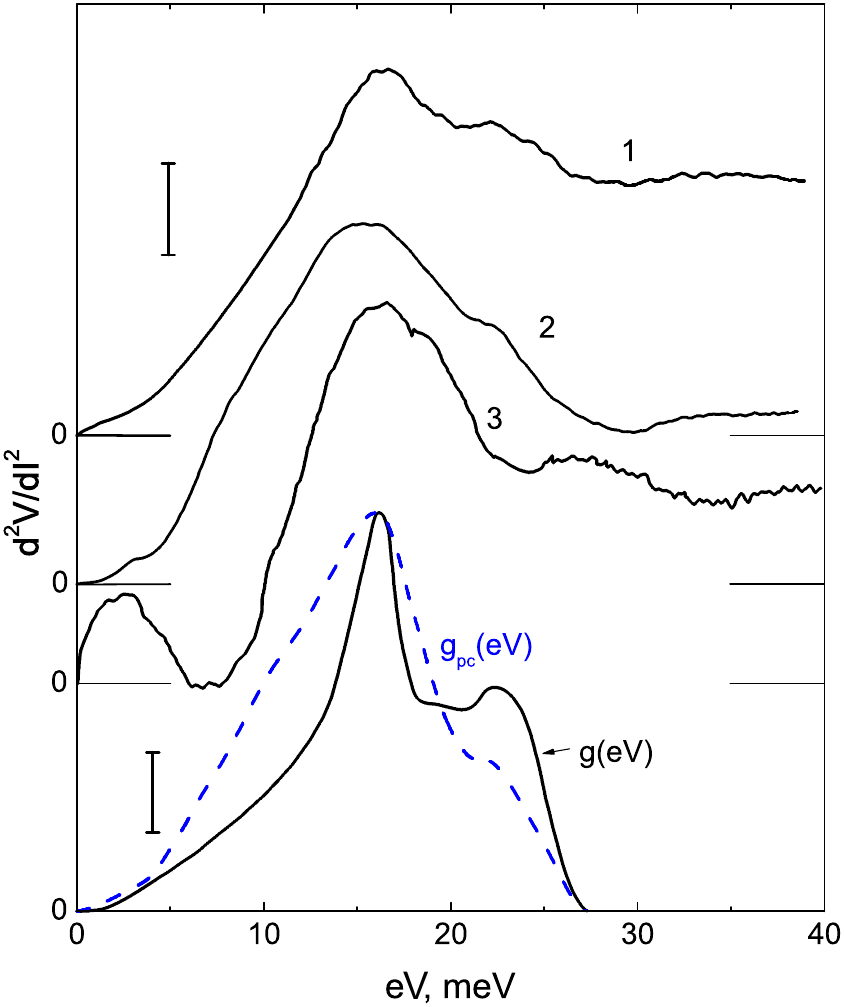}
\caption[]{Point-contact spectra of niobium in the normal state at $T=4.2 K$, $H=40-50\ kOe$ (curves 1-3). The resistances of the contacts, the values of the modulating voltage with $V=0$, and the value of the vertical segment with the horizontal bars, indicating the calibration for the second harmonic ${{V}_{2}}\sim d{{V}^{2}}/d{{I}^{2}}$, are equal to 110, 334, and 185 $\Omega $ ; 0.993, 2.26, and 2.47 $mV$; 0.68, 0.555, and 0.238 $\mu V$ for curves 1, 2, and 3, respectively (${{g}_{pc}}$  is the point contact EPI function, obtained from spectrum 1 in the model of a clean orifice \cite{Khotkevich}; $g$ is the tunneling EPI function from Ref. \cite{Robinson}; the value of the vertical segment with the horizontal bars equals 0.023 and 0.112 dimensionless units for ${{g}_{pc}}$ and $g$, respectively).}
\label{Fig1}
\end{figure}
It is very difficult to obtain point contacts from niobium corresponding to the model of a clean orifice and satisfying the conditions of ballistic flight of electrons. The point-contact EPI spectra for specimens satisfying these conditions have maximum intensity, low background level, and quite narrow spectral lines. It is evident that even the best niobium spectra that we obtained and which are shown in Fig. \ref{Fig1} (curve 1) do not completely satisfy these requirements. Let us compare the point contact spectrum 1 and the point contact EPI function ${{g}_{pc}}(\omega )$ obtained from it by subtracting out the background with the function $g(\omega )$, measured by the tunneling method in Ref. \cite{Robinson}. The position of the maxima on the curves being compared coincide, but the smooth maximum on the point-contact spectrum is smeared, which could possibly indicate the inhomogeneity of the structure of the metal in the region of the constriction. If the parameter ${{\lambda }_{pc}}=2\int\limits_{0}^{\infty }{{{g}_{pc}}(\omega )d\omega /\omega }$ is calculated for contact 1 using equations for the clean orifice, then we shall obtain a value close in order of magnitude to the standard EPI parameter $\lambda =0.82$ \cite{Robinson,Wolf}. The absolute values of the function ${{g}_{pc}}(\omega )$ and of the EPI parameter ${{\lambda }_{pc}}$  depend on the choice of $\rho l={{p}_{F}}/n{{e}^{2}}$ and ${{v}_{F}}$, characterizing $Nb$ in the free electron model. In Fig. \ref{Fig1}, the calibration of the ordinate scale for ${{g}_{pc}}(\omega )$ is presented for $\rho l=0.37\cdot {{10}^{-11}}\ \Omega \cdot c{{m}^{2}}$  and ${{v}_{F}}=0.294\cdot {{10}^{8}}\,cm/\sec $ (Ref. \cite{Khotkevich}) here; ${{\lambda }_{pc}}=0.23$. There values of $\rho l$ and ${{v}_{F}}$  will be used in the estimates of different parameters in the paper. If the value $\rho l=0.47\cdot {{10}^{-11}}\ \Omega \cdot c{{m}^{2}}$ and ${{v}_{F}}=0.897\cdot {{10}^{8}}\,cm/\sec $ taken from Ref. \cite{Blonder} are substituted, then for contact 1 we obtain ${{\lambda }_{pc}}=0.62$. It is interesting to note that many point-contact spectra have a shoulder at $\approx 10\text{ }meV$ and a break in the region 3-4 $meV$, which are also present on some tunneling EPI spectra of niobium (see, for example, Ref. \cite{Wolf}). Since the inelastic mean-free path of electrons ${{l}_{\varepsilon }}={{v}_{F}}{{\tau }_{\varepsilon}}$  or the diffusion length for energy relaxation in a dirty material ${{\Lambda }_{\varepsilon }}{{({{l}_{i}}{{l}_{\varepsilon }}/3)}^{{1}/{2}\;}}$ (${{l}_{i}}\ll {{l}_{\varepsilon }},\ \,{{l}_{i}}$  is the momentum free path length) decrease as the energy increases, it is natural to expect that the point contact better satisfies the conditions of ballistic or diffusion regimes in the low-voltage range. Starting from the tunneling EPI functions, it is possible to calculate the energy relaxation time $\tau _{\varepsilon }^{-1}=2\pi \int\limits_{0}^{\infty }{g(\omega )d\omega}$  at $T=0$. For electrons with energies close to the maximum phonon energy, we obtain ${{l}_{\varepsilon }}=60\ {\AA}$ , which exceeds the diameter of the contact 1, determined using the equation for the clean (${{l}_{i}}\gg d$) orifice $d=1.3{{(\rho l/R)}^{1/2}}24\ {\AA}$ . Curves 2 and 3 in Fig. \ref{Fig1} represent point-contact spectra of specimens for which a short momentum free path length $l_i$ is apparently characteristic. This is indicated by the following factors. First, the intensities of the point-contact spectra are such that, having been analyzed using the equations for the model of a clean orifice, they lead to considerably lower values of ${{\lambda}_{pc}}$  (0.08 and 0.02 for contacts 2 and 3, respectively) than for spectrum 1. Second, the spectral bands are either strongly smeared, as occurs for curve 2, or do not coincide with the positions of the maxima in the tunnel EPI function. The last remark refers, in particular, to the second maximum on curve 3, which is displaced toward higher energies by $\approx 5\text{ }meV$ (its position corresponds to $eV\approx 27\text{ }meV$) compared with $22\ meV$, as for the function $g(\omega )$  or curve 1. We also observed intermediate positions of the second maximum, for example, at $eV\approx 25\text{ }meV$), for some contacts. In some cases the position of the first maximum remained practically unchanged ($eV\cong 16\text{ }meV$), which eliminates systematic errors in the measurements as a possible source of the displacements noted. Apparently, the "creeping" of the second maximum along the energy axis is due to inhomogeneities of the structure and the composition of the material filling the region of the constriction and including conducting niobium oxides in the form of impurities.

The presence of distinct spectral maxima in curves 1-3 (Fig. \ref{Fig1}) indicates that at least the diffusion length for energy relaxation is much greater than the dimensions of the contacts. It is impossible to estimate quantitatively the mean-free path length and the diameter of contacts 2 and 3, since the electronic parameters of the material forming the constriction are unknown. Unfortunately, we were not able to measure the point-contact spectra in the superconducting state for contacts which had spectra such as curves 1-3 in Fig. \ref{Fig1} in the normal state. Most contacts generally did not have distinct spectral features in the second derivatives of the IVC in the normal state. The results of the investigation of point-contact spectra for these specimens in the superconducting state are presented in the next section.
\subsection{Point-contact spectra of niobium in the superconducting state}
From the large variety of IVC of niobium contacts in the superconducting state, we selected for analysis only the characteristics, for which the effect of Joule heating could be neglected. The voltage dependence of the excess current served as a criterion. The excess current was defined as the difference between the nonlinear IVC in the normal and superconducting states. Thermal effects were assumed to be insignificant if the excess current decreases with increasing voltage by an amount that is a small fraction (of the order of several percent) of its initial value, reached in the region $eV\gtrsim 2\Delta $ (${{\Delta }_{Nb}}=1.4\ meV$). Below, we present more direct proof of the insignificance of thermal effects for the contacts being discussed.

\begin{figure}[]
\includegraphics[width=8cm,angle=0]{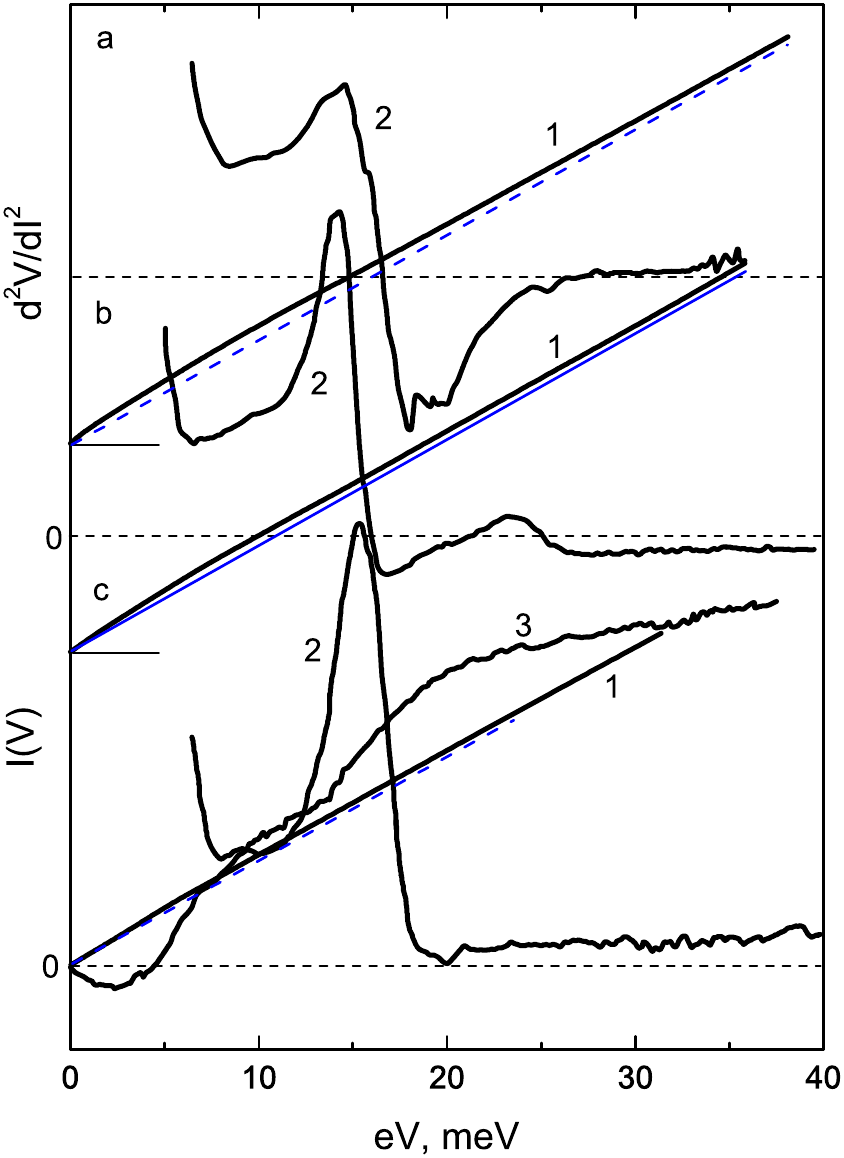}
\caption[]{Current-voltage characteristics (curve 1) and second derivatives (curve 2), for dirty superconducting niobium contacts ($T=4.2-4.5\ K, H=0$). The resistances of contacts $a, b$, and $c$ are 194, 17.5, and 29.2 $\Omega $, respectively. Curves 1a and 3 describe the IVC and its second derivative for contacts $b$ and $c$ in the normal state (at $T=10\ K$). The modulating voltages for curves 2 and 3 in Fig. \ref{Fig2}c equal 0.761 and 2.26~$mV$. The contacts $b$ and $c$ contain aluminum oxide.}
\label{Fig2}
\end{figure}

Figure \ref{Fig2}
shows typical IVC (curves 1) and their second derivatives (curves 2), measured in zero field at a temperature of $\approx 4.2\ K$  (niobium in the superconducting state). The second derivatives have features for bias voltages $eV$ close to the energy of the main peak in the EPI function of niobium $\hbar {{\omega }_{TA}}=16\ meV$, owing to scattering of electrons by transversely polarized phonons (see Fig. \ref{Fig1}). In many spectra, a weaker feature could also be observed at $eV=22-24\ meV$ (see, e.g., curve 2 in Fig. \ref{Fig2}b), due to longitudinal phonons. Current-voltage characteristics and point-contact spectra in the normal state (for $T\gtrsim {{T}_{c}}$) (see curve 1 in Fig. \ref{Fig2}b and curve 3 in Fig. \ref{Fig2}c) were measured for a number of contacts. The form of the point-contact spectra in the normal state is analogous to curve 3 in Fig. \ref{Fig2}c for other specimens as well, and the spectral maxima in the second derivatives are absent in the normal state. During the course of measurements, under the influence of external electric adjustments, the resistance of many contacts changed irreversibly and the form of the IVC and the point-contact spectra also changed with it, which did not permit obtaining a complete set of characteristics in the superconducting and normal states for all specimens. For contacts whose spectra are shown in Figs. \ref{Fig2}a and c, the IVC in the normal state were not obtained, so that we present their presumed form with the dashed lines, passing through the origin of coordinates and parallel to the excess current in the region of voltages 5-16 $mV$, i.e., before the main maximum in the second derivatives. The magnitude of the excess current in most cases was smaller than the theoretical value for dirty $S-c-S$ contacts \cite{Artemenko}  $I_{exc}^{(0)}=({{\pi }^{2}}/4-1)(\Delta /eR)$. This can be interpreted as being a result of the elastic scattering of electrons by impurities or defects in the region of the constriction \cite{Blonder}. The smallness of the mean free path $l_i$ compared to the size of the contact also indicates the absence of a large number of subharmonics of the gap which, according to Ref. \cite{Klapwijk}, result from repeated passes of normal excitations, undergoing Andreev reflection from the $N-S$ boundaries, through the region of the constriction. We usually observed the smeared maximum in the second derivative for bias voltages of $eV\approx \Delta \div2\Delta $  and not more than one additional maximum for lower voltages, which could be interpreted as the manifestation of second-order harmonics.

The absence of spectral bands in the second derivatives of the IVC in the normal state indicates the smallness of the energy relaxation length ${{\Lambda }_{\varepsilon }}$  compared with the characteristic size of the contact $d$. In this case, we assume that the elastic mean-free path ${{l}_{i}}$  is much shorter than the inelastic length ${{l}_{\varepsilon }}$. The coherence length in the region of the constriction equals $\xi \approx {{({{\xi }_{0}}{{l}_{i}})}^{1/2}}\ ({{\xi }_{0}}\cong 400 {\AA})$, since in the limitingly dirty contacts the inequality ${{\xi }_{0}}\gg {{l}_{i}}$ is satisfied. Therefore, in the contacts that we are examining, the conditions
\begin{equation}
\label{eq__2}
{{{\xi }}\gtrsim d>{{\Lambda }_{\varepsilon }}}
\end{equation}
which lead to the spectral features observed in Fig. \ref{Fig2} on the derivatives of the IVC in the superconducting state, are satisfied.

The form of the IVC at low voltages ($eV\lesssim \Delta $) permits assuming that in many cases, real niobium point contacts have a much more complicated structure than an $S-c-S$ contact with a homogeneous composition. This is indicated, first of all, by the absence of any traces of a critical current and, second, by the proximity of the differential resistance at $V=0$ in the superconducting state to its magnitude in the normal state (these two quantities differ, as a rule, by not more than a factor of two). Due to the inhomogeneous distribution of elastic scatterers of electrons and centers which break up Cooper pairs, in the region of the constriction the coherence length also depends on the coordinates, assuming a minimum value at the center of the contact and increasing with distance into the interior of the edges.
\begin{figure}[]
\includegraphics[width=8cm,angle=0]{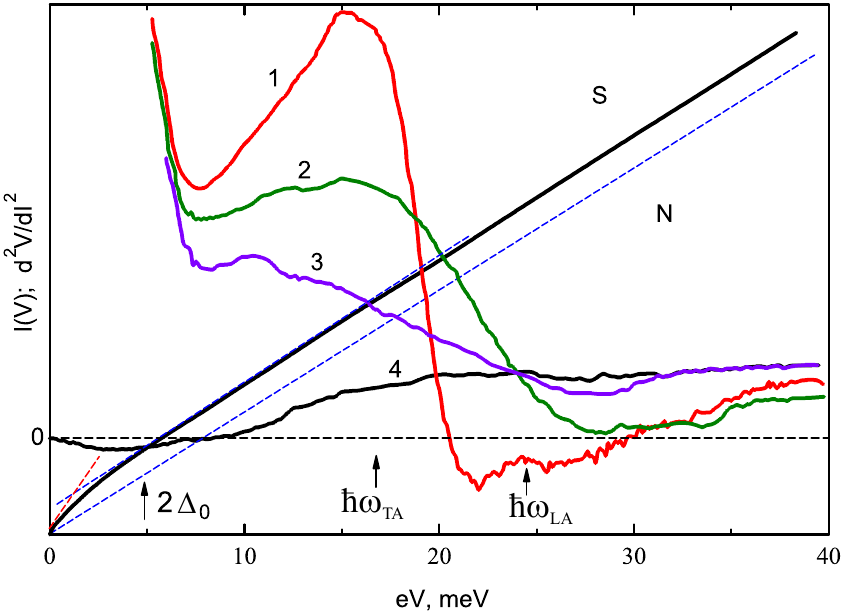}
\caption[]{Point-contact spectra of the same contact obtained at different temperatures in zero magnetic field. Curves 1, 2, 3, and 4 correspond to temperatures 4.5, 6.5, 8.65, and 10.6 $K$. The IVC in the superconducting ($S$) and normal ($N$) states are measured at temperatures of 7.8 and 11.2 $K$, respectively. The resistance of the contact is 30 $\Omega$. The niobium electrodes are covered with an aluminum oxide.}
\label{Fig3}
\end{figure}

Figure \ref{Fig3} shows the second derivatives of IVC measured at temperatures both lower and higher than ${{T}_{c}}$  for niobium. As ${{T}_{c}}$ is approached from below, the spectral feature near $\hbar {{\omega }_{TA}}$  becomes wider and its intensity drops, but the position of the $V$ axis is \emph{not displaced} toward lower voltages, as would be expected if the observed effect were due to heating or, for example, if the current density were close to the critical value. Therefore, the position of the features that we observed is determined by the voltage drop across the contact. It is also evident that the observed nonlinearity in the superconducting state is entirely due to the energy dependence of the excess current and is not related to the nonlinearity of the IVC in the normal state.
\subsection{Comparison with theory for the clean limit}
There is no theory for the energy dependence of the excess current in dirty superconducting contacts. However, the nature of the appearance of the excess current on IVC for clean and dirty contacts is the same. For this reason, it is interesting to compare the form of the observed features on the second derivative of the IVC of dirty contacts with the theoretically predicted form of the spectral features for clean contacts. This can be done most completely for $S-c-N$ contacts.
\begin{figure}[]
\includegraphics[width=8cm,angle=0]{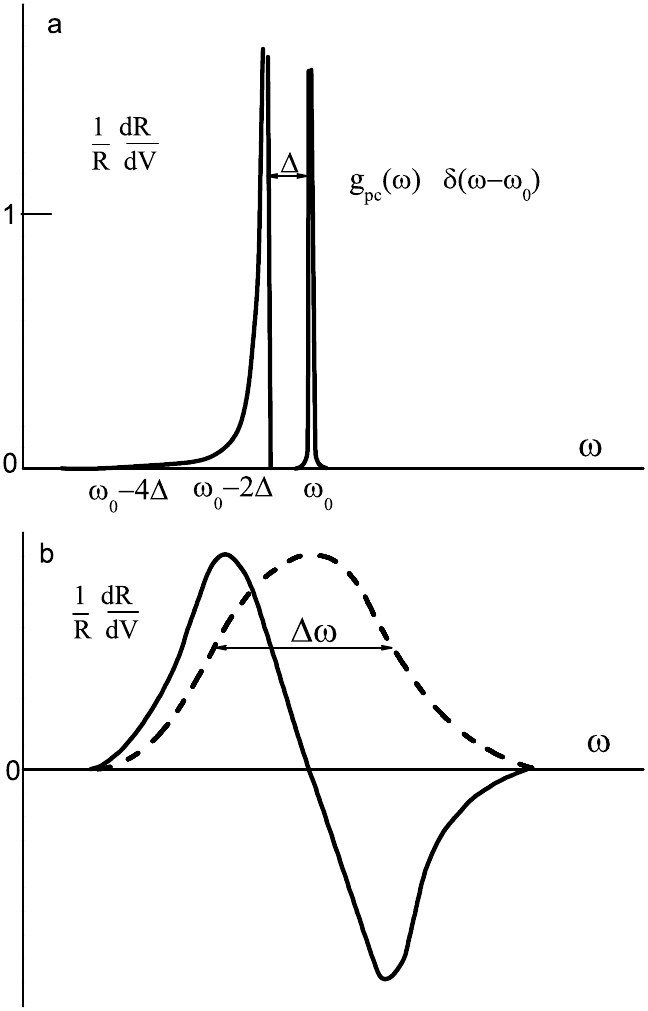}
\caption[]{Theoretical form of the features in the second derivative of the excess current with respect to the voltage far clean $S-c-N$ contacts: a) for an infinitely narrow line in the EPI function; b) for a band whose width is much greater than $\Delta $.}
\label{Fig4}
\end{figure}
According to Ref. \cite{Yanson}, the form of the line depends on the ratio between the width of the peak in the EPI spectrum and $\Delta $. Let a narrow line at the frequency ${{\omega }_{0}}$, which can be approximated by a $\delta $-function, shown in Fig. \ref{Fig4}a by the dashed line, be present in the phonon spectrum. Then, the second derivative of the IVC $\left( 1/R \right)\left( dR/dV \right)$,  which is proportional to ${{d}^{2}}V/d{{I}^{2}}$ measured in the experiment, when the nonlinearity is small, has the form
\begin{equation}
\label{eq__3}
{\frac{1}{R}\frac{dR}{dV}\sim \theta (x-1)\frac{2{{\left( x-\sqrt{{{x}^{2}}-1} \right)}^{2}}}{\sqrt{{{x}^{2}}-1}}}
\end{equation}
for $T\ll \Delta $, where $x=(\hbar {{\omega }_{0}}-eV)/\Delta $. This function is illustrated in Fig. \ref{Fig4}a by the continuous line and has the form of a sharp peak, situated at $eV=\hbar {{\omega }_{0}}-\Delta$. For the reverse relation between the width of the line $\Delta \omega $  and $\Delta $, it is the first and not the second derivative that is approximately proportional to the EPI function and, therefore, the second derivative has the form of the curve shown by the continuous line in Fig. \ref{Fig4}b. Comparing the point-contact spectra in the superconducting state (curves 1 in Fig. \ref{Fig2}) with the theoretical curves presented in Fig. \ref{Fig4}, it is evident that both limiting (Fig. \ref{Fig2}a and c) and intermediate cases (Fig. \ref{Fig2}b) are observed. The specific form of the curve observed experimentally apparently depends on the composition of the material in the region of the constriction, from which information on EPI is obtained within the framework of PCS.

As noted above, the nonlinear features due to the energy dependence of the excess current, according to the theory of clean $S-c-N$ or $S-c-S$ contacts, constitute a small (of the order of $\Delta /eV$) fraction of the nonlinear- ities owing to the non-ohmic nature of IVC in the normal state and leading, as is well known \cite{Yanson}, to a proportionality between its second derivative and the EPI function. As is evident from Figs. \ref{Fig2} and \ref{Fig3}, in dirty contacts, practically the entire observed nonlinearity is due to the function $I_{S}^{(1)}(eV)$, while the nonlinearity in the normal state is only a small part of it. For example, the increase in the differential resistance in the interval from 0 to 30 $meV$ constitutes, in the superconducting state, usually 4-5\% of $R_0$, while in the normal state, for the same contacts, it constitutes only 1\%. This is clearly shown in Fig. \ref{Fig3}, where the point-contact spectra in the superconducting (curves 1-3) and normal (curve 4) states are presented on a single scale. We note that although curve 3 in Fig. \ref{Fig2}c is also presented on the same scale with curve 2 (the voltage of the second harmonic $V_2$, proportional to ${{d}^{2}}V/d{{I}^{2}}$, is plotted along the ordinate axis), it is obtained with a considerably higher level of modulating voltage $V_1$, to whose square $V_2$ is proportional.

As for the reason why PCS is possible in limitingly strong contacts, we note the following. Generally, phonon spectroscopy in point contacts is possible due to the fact that two groups of electrons, whose maximum energies differ by $eV$, coexist in some region of the metal. In the normal state in dirty ($d\gg {{\Lambda }_{\varepsilon }}$) contacts, this condition is not satisfied and the spectral lines do not appear in the second derivative. However, with the transition into the superconducting state, it may again turn out that in a region much smaller than ${{\Lambda }_{\varepsilon }}$, two groups of electrons, whose energies differ by $eV$ will coexist. These electrons arise as a result of the breakup of Cooper pairs in the region of dimensions of the order of ${{\Lambda }_{\varepsilon }}$ near the center of the contact, where there is a jump in the pair chemical potential. The chemical potentials of pairs are constant to the left and right of the jump and the difference between them equals $eV$, in spite of the presence of the electric field and the smooth behavior of the chemical potential of quasi-particles on a scale of the order of $d$.

There is one more question that must be resolved in examining the properties of limitingly dirty contacts. This concerns the obvious fact that the conditions of the thermal limit are not satisfied \cite{Verkin}, in spite of the fact that $l_i$ and ${{\Lambda }_{\varepsilon }}$. are much smaller than $d$. According to the theory of the thermal limit, the temperature at the center of the contact with $eV=16\ meV$ must reach 51 $K$, at which the excess current must definitely be absent. The inapplicability of the thermal limit in this case is apparently due to the fact that the electronic thermal conductivity constitutes only a part of the total thermal conductivity of the dirty metal, while the phonon mean-free path is greater than the dimensions of the contact. An analogous situation apparently occurs also in the point-contact spectroscopy of resistive alloys and compounds, in particular compounds with fluctuating valence \cite{Bussian}.
\section{CONCLUSIONS}
In this paper, we have shown that point-contact phonon spectroscopy is possible in small limitingly dirty superconducting contacts, in which Joule heating can be neglected. In addition, there are no peaks in the second derivative of the IVC in the normal state. Contacts with the required characteristics can be prepared from niobium. The results obtained give hope that the EPI function of superconducting compounds with high critical parameters can also be measured using this method. Further investigations will show the extent to which this technique can be applied to other $d-$ and $f-$ superconducting metals and their alloys.
\section{NOTATION}
Here $l_i$ is the electron momentum mean-free path; ${{l}_{\varepsilon }}$ is the inelastic mean-free path of electrons; ${{\Lambda }_{\varepsilon }}$ is the diffusion length for electron energy relaxation; $\xi$  is the coherence length; $d$ is the constant diameter of the constriction; $\Delta $ is the energy gap; ${{\tau }_{\varepsilon }}$  is the energy relaxation time; $v_F$ and $p_F$ are the Fermi velocity and momentum; $g$ is the tunneling electron-phonon interaction function; $g_{pc}$ is the point-contact electron-phonon interaction function; $V$ is the voltage drop across the contact; and $I$ is the current through the contact.

\end{document}